\documentclass[12pt,a4paper]{article}
\usepackage{amssymb}
\usepackage{graphicx}
\usepackage{amsmath}

\begin{document}

\title{Trajectory trapping and the evolution of drift turbulence beyond the
quasilinear stage}
\author{{\small Madalina Vlad} \\
%EndAName
{\small National Institute for Laser, Plasma and Radiation Physics, }\\
{\small P.O.Box MG-36, Magurele, Bucharest, Romania}}
\maketitle

\begin{abstract}
Test modes on turbulent magnetized plasmas are studied taking into account
the ion trapping that characterizes the $\mathbf{E}\times \mathbf{B}$ drift
in the background turbulence. We show that trappyng provides the physical
mechanism for the formation of large scale potential structures (inverse
cascade) observed in drift turbulence. Trapping combined with the motion of
the potential with the diamagnetic velocity determines ion flows in opposite
directions, which reduce the growth rate and eventually damps the drift
modes. It also determine transitory zonal flow modes in connection with
compressibility effect due to the polarization drift in the background
turbulence.


\vspace*{0.30in}
\end{abstract}

\vspace*{0.5cm}

%\newline

\section{Introduction}

The evolution of turbulence in magnetically confined plasmas is a complex
problem that is not yet completely understood besides the huge amount of
work on this topic (\cite{K02} and the references there in). Low-frequency
drift type turbulence, which has significant influence on the magnetic
confinement of high temperature plasmas, is extensively studied especially
in connection with fusion research (see e.g. \cite{Horton}, \cite{garbet}, 
\cite{tynan}). Most of the studies that go beyond the quasilinear stage are
based on numerical simulations or on simplified models. They show a complex
nonlinear evolution with generation of large scale structures, increase of
order and appearance of zonal flow modes (\cite{Tzf}, \cite{Diamond}) that
leads to the nonlinear damping of turbulence.

The stochastic particle advection that appears in turbulent plasmas due to
the $\mathbf{E}\times \mathbf{B}$ (or electric) drift determined by the
fluctuating potential can produce trajectory trapping or eddying around
contour lines of the potential. Trapping has a strong effect on the
statistics of particle trajectories. Analytical methods adequate to the
study of particle stochastic advection in the presence of trapping were
developed only in the last decade (\cite{V98}, \cite{VS04} and the
references therein). They permitted to understand that trapping determines
strong departure from the characteristics of Gaussian advection processes.
The aim of this paper is to contribute to the understanding of the effects
of trajectory trapping on the evolution of drift turbulence. These are the
first analytical results on this complex problem that are in agreement with
numerical simulations.

Test particle trajectories are strongly related to plasma turbulence. Plasma
dynamics is basically described by the Vlasov-Maxwell system of equations,
which represents the conservation laws for the electron and ion distribution
functions along particle trajectories coupled with the constraints imposed
by Maxwell equations. Analytical studies of plasma turbulence based on
trajectories were initiated by Dupree \cite{D66}, \cite{D72} and developed
especially in the years seventies \cite{K02}. These methods do not account
for trajectory trapping and thus they apply to the quasilinear regime or to
unmagnetized plasmas. A very important problem that has to be understood is
the effect of the non-standard statistical characteristics of the test
particle trajectories on the evolution of turbulence in magnetized plasmas.
A Lagrangian approach is developed, which extends the Lagrangian methods of
the type of \cite{D72}, \cite{SV88}, \cite{VSM} to the nonlinear regime
characterized by trapping.

We study linear modes on turbulent plasma with the statistical
characteristics of the potential considered known. Drift turbulence in hot
magnetized plasmas is considered. Analytical expressions are derived, which
approximate the growth rates and the frequencies of the test modes as
functions of the characteristics of the background turbulence. They provide
an image of turbulence evolution.

We show that there is a sequence of processes, which appear at different
stages of evolution as transitory effects and that the drift turbulence has
an oscillatory (intermittent) evolution. A different perspective on
important aspects of the physics of drift type turbulence in the strongly
non-linear regime is deduced. The main role in these processes is shown to
be played by ion trapping.

The paper is organized as follows. The dispersion relation for the drift
modes on turbulent plasmas is deduced in Section 2. It is shown that the
effects of the background turbulence are contained in a function of time
that enters in the ion propagator. It is an average over the stochastic ion
trajectories in the background turbulence. The statistical methods for
evaluating this function are shortly presented in Section 3. The next three
Sections present the effects of the turbulence on the test modes for
quasilinear turbulence, weak nonlinear regime (when the fraction of trapped
ions is small) and strong nonlinear regime (with strong trapping). The
evolution of the drift turbulence is discussed at each stage. The summary of
results and the conclusions are presented in Section 8.

\section{Test modes in turbulent plasmas}

Drift waves and instabilities are low-frequency modes generated in
non-uniform magnetically confined plasmas. Depending on the particular
conditions, there are several types of drift modes. Since the aim of this
work is to understand the effects of trapping on the evolution of
turbulence, we consider a simple confining geometry, the plane plasma slab,
in which the magnetic field is straight and uniform. Plasma has low $\beta $%
, which means that the perturbation of the magnetic field is negligible
(electrostatic approximation).

The magnetic field is along $z$ axis ($\mathbf{B}=B\mathbf{e}_{z})$ and
plasma is non-uniform in one direction taken along $x$ axis. For simplicity,
the equilibrium temperatures are uniform and only the density $n_{0}(x)$ is $%
x$-dependent. The characteristic length of density variation $%
L_{n}=n_{0}/\left\vert dn_{0}/dx\right\vert $ is much larger than the wave
length of the drift modes.

We start from the basic description of this (universal) drift turbulence
provided by the drift kinetic equation in the collisionless limit. Electron
kinetic effects produce the dissipation mechanism to release the energy and,
combined with finite Larmor of the ions, make drift waves unstable. The
latter consist of the polarization drift velocity and of the modification of
the electric drift velocity due to the gyro-average of the potential on the
ion orbits. Both effects determine the decrease of mode frequency below the
diamagnetic frequency, which makes the growth rate positive. Beside this,
the polarization drift has a more complex influence determined by its
nonzero divergence. We neglect here the modification of the potential and
consider the effects of the polarization drift. The reason is that, as shown
below, the background turbulence in the nonlinear regime eliminates the
effects of the finite Larmor radius on the frequency. Thus, this
approximation that determines sensible simplifications of the calculations
has negligible influence on the phenomenology of \ drift turbulence
evolution in the nonlinear regime.

The drift kinetic equations for the perturbation $\delta f^{\alpha }$ of the
distribution function of the guiding centers $f^{\alpha }(\mathbf{x}%
,z,v_{z},t)=F^{\alpha }(v_{z})\ n_{0}(x)+\delta f^{\alpha }(\mathbf{x}%
,z,v_{z},t)$ for electrons and ions is 
\begin{equation}
\partial _{t}\delta f^{\alpha }+\mathbf{u}_{\perp }\cdot \mathbf{\nabla }%
\delta f^{\alpha }+v_{z}\partial _{z}\delta f^{\alpha }-\frac{e_{\alpha }}{%
m_{\alpha }}\left( \partial _{z}\phi \right) \partial _{v_{z}}\delta
f^{\alpha }=F^{\alpha }(v_{z})\ \partial _{y}\phi \ \partial _{x}n_{0}
\label{gk}
\end{equation}%
where $\alpha $\ represents the species ($\alpha =e,i),$ $\phi $ is the
potential, $\mathbf{x}=\left( x,y\right) ,$ $F^{\alpha }(v_{z})$ is the
Maxwell distribution,$\ v_{z}$\ is the velocity along the magnetic field and 
$e_{\alpha },$ $m_{\alpha }$ are the charge and mass of the particles. The
perpendicular velocity is $\mathbf{u}_{\perp }=-\mathbf{\nabla }\phi \times 
\mathbf{e}_{z}/B.$

The perturbed densities $\delta n^{\alpha }$ are obtained by integrating the
distribution function $\delta f^{\alpha }$ over the velocities $v_{z}$ and $%
\mathbf{v}_{\perp }$. Poisson equations, which is\ approximated by the
quasi-neutrality condition $\delta n^{e}=\delta n^{i},$ closes the system.

\bigskip

The instabilities are usually studied on quiescent plasmas by introducing in
Eq. (\ref{gk}) a wave type potential $\phi =\delta \phi $ where%
\begin{equation}
\delta \phi (x,y,z,t)=\phi _{k\omega }\exp \left(
ik_{x}x+ik_{y}y+ik_{z}z-i\omega t\right) .  \label{fi}
\end{equation}%
Linearizing the equations, the frequency $\omega $ and the growth rate $%
\gamma $ are determined from the quasineutrality condition, which is the
dispersion relation for the drift waves.

This model of modes developing on quiescent plasma is not realistic because
drift instabilities appear for a large range of wave numbers and produce a
turbulent potential. Test mode models consider a turbulent plasma with given
statistical characteristics of the background potential $\phi _{b}(\mathbf{x}%
,z,t)$ and a small perturbation $\delta \phi ,~\phi =\phi _{b}+\delta \phi .$
The growth rates and the frequencies of the test modes are determined as
functions of the statistical characteristics of the background potential $%
\phi _{b}$ by linearizing Eq. (\ref{gk}) around the background potential.

The aim of the present study is to determine the effects of the turbulence
on test modes and in particular the influence of trajectory trapping or
eddying produced by the electric drift. The dispersion relation of the drift
waves in quiescent plasma ($\phi _{b}=0)$ is review in subsection 2.1 in
order to have a comparison basis for the effects appearing in turbulent
plasmas. The dispersion relation for turbulent plasmas is determined in
subsection 2.2, where we show that the background potential modifies the
propagator of the modes through trajectory distribution and also by a
compressibility effect produced by the polarization drift.

\subsection{Drift modes in quiescent plasmas}

We begin with the electrons and show that their response is the same in
quiescent and turbulent plasmas because actually they do not "see" the
turbulence due to the fast decorrelation produced by the motion along the
magnetic field.

Electrons are dominated by the parallel motion for the small frequency
fluctuations with $\omega \ll k_{z}v_{Te}.$ The density of electrons can be
written as 
\begin{equation}
f^{e}\left( \mathbf{x},z,\mathbf{v},t\right) =n_{0}(x)F_{M}^{\alpha }\exp (%
\frac{e\phi }{T_{e}})+h^{e}  \label{re}
\end{equation}%
where the first term is the adiabatic response obtained from the parallel
terms in Eq. (\ref{gk}) and the non-adiabatic term $h^{e}$ is the solution
of 
\begin{eqnarray}
&&\partial _{t}h^{e}-\frac{\mathbf{\nabla }\phi \times \mathbf{e}_{z}}{B}%
\cdot \mathbf{\nabla }h^{e}+v_{z}\partial _{z}h^{e}-\frac{e_{\alpha }}{%
m_{\alpha }}\left( \partial _{z}\phi \right) \partial _{v_{z}}\delta
f^{\alpha }  \label{gke} \\
&=&n_{0}F_{M}^{e}\left( V_{\ast e}\partial _{y}+\partial _{t}\right) \frac{%
e\phi }{T_{e}},  \notag
\end{eqnarray}%
where the diamagnetic velocity%
\begin{equation*}
V_{\ast e}=-\frac{T_{e}}{en_{0}B}\frac{\partial n_{0}}{\partial x_{1}}=c_{s}%
\frac{\rho _{s}}{L_{n}}
\end{equation*}
was introduced and the condition $e\phi /T_{e}<1$ that holds for the drift
waves was used. $T_{e}$ is the electron temperature, $c_{s}=\sqrt{T_{e}/mi},$
$\rho _{s}=c_{s}/\Omega _{i}$ is the Larmor radius and $\Omega _{i}=eB/m_{i}$
is ion cyclotron frequency.

Due to the large electron velocity $v_{z},$ the perpendicular drift can be
neglected in electron trajectories as well as the parallel acceleration
which produces velocity variations that are negligible compared to the
thermal velocity. The trajectories are not influenced by the potential and
consequently electron equation for drift waves can be approximated by the
linear one%
\begin{equation}
\partial _{t}h^{e}+v_{z}\partial _{z}h^{e}=n_{0}F_{M}^{e}\left( V_{\ast
e}\partial _{y}+\partial _{t}\right) \frac{e\phi }{T_{e}}.  \label{lgke}
\end{equation}%
The solution is obtained by using the method of characteristics. The right
hand side of Eq. (\ref{lgke}) is integrated along the trajectory, which is
approximated by $z(\tau )=z-v_{z}(t-\tau ).$ For the potential (\ref{fi}) it
is%
\begin{equation}
h^{e}=\frac{e\phi _{k\omega }}{T_{e}}n_{0}F_{M}^{e}\frac{\omega
-k_{y}V_{\ast e}}{v_{z}k_{z}-\omega }\exp \left( i\mathbf{k}.\mathbf{x}%
+ik_{z}z-i\omega t\right) .  \label{he}
\end{equation}%
The perturbation of the electron density is obtained by integrating $f^{e}$
over velocities. The integral aver $v_{z},$ which is singular, is determined
in the complex $\omega $ plane by the pole of \ the singularity (the
principal value is negligible) [see, for example, \cite{GR}, page 457]. One
obtains 
\begin{equation}
\delta n^{e}=n_{0}(x)\frac{e\phi _{k\omega }}{T_{e}}\left( 1+i\sqrt{\frac{%
\pi }{2}}\frac{\omega -k_{y}V_{\ast e}}{\left\vert k_{z}\right\vert v_{Te}}%
\right) \exp \left( i\mathbf{k}.\mathbf{x}+ik_{z}z-i\omega t\right) ,
\label{edn}
\end{equation}%
which hold for both quiescent and turbulent plasmas.

\bigskip

The ion response is influenced by finite Larmor radius effects: the
polarization drift determined by the time variation of the electric field
and the modification of the potential by the gyro-average. These effects
combined with the non-adiabatic response of the electrons destabilize the
drift waves. In order to concentrate on the nonlinear effects produced by
trajectory trapping, we make an additional simplification, which has the
advantage of simplifying the analytical expressions: we neglect the change
of the potential by the gyro-average and take the same potential in the ion
and electron equations. The reason is that the turbulence evolves in the
nonlinear stage to correlation lengths much larger than the ion Larmor
radius and the effect of potential gyro-average becomes negligible. As we
are interested in this paper by effects that appear beyond the quasilinear
stage, this simplification does not affect the results.

The polarization drift 
\begin{equation}
\mathbf{u}_{p}=\frac{m_{i}}{eB^{2}}\partial _{t}\mathbf{E}_{\perp }
\label{pol}
\end{equation}%
is much smaller than the electric\ drift (by a factor $\omega /\Omega _{i},$
where) and actually it has negligible effect on ion trajectories. The
polarization drift is important in the ion equation due to its divergence%
\begin{equation}
\mathbf{\nabla }_{\perp }\cdot \mathbf{u}_{p}=-\frac{m_{i}}{eB^{2}}\partial
_{t}\Delta \phi ,  \label{dpol}
\end{equation}%
which leads to the perturbation of the ion density. We consider perturbation
with small parallel wave numbers $\omega /k_{z}\gg v_{Ti}$ and the parallel
motion can be neglected in Eq. (\ref{gk}) for the ions. The linearized
equation with the potential (\ref{fi}) is 
\begin{equation}
\partial _{t}\delta f^{i}=-n_{0}\frac{eF_{M}^{i}}{T_{e}}V_{\ast e}\partial
_{y}\delta \phi -n_{0}F_{M}^{i}\mathbf{\nabla \cdot u}_{p}  \label{gki}
\end{equation}%
The solution is%
\begin{equation}
\delta f^{i}(\mathbf{x},t,\mathbf{v)}=-n_{0}\frac{eF_{M}^{i}}{T_{e}}\delta
\phi \left[ k_{y}V_{\ast e}-\frac{T_{e}m_{i}}{e^{2}B^{2}}\omega k_{\perp
}^{2}\right] \Pi ^{i}  \label{dfi}
\end{equation}%
where $k_{\perp }=\sqrt{k_{x}^{2}+k_{y}^{2}}.$\ The propagator $\Pi ^{i}$ is
defined by 
\begin{equation}
\Pi ^{i}=i\int_{-\infty }^{t}d\tau \exp \left( i\mathbf{k\cdot }\left( 
\mathbf{x}(\tau )-\mathbf{x}\right) (\tau )-i\omega (\tau -t)\right)
\label{prop}
\end{equation}%
with the integral taken along ion trajectories. In the linearized case $\Pi
^{i}=-1/\omega $. The perturbation of the ion density is obtained by
integrating over velocities%
\begin{equation}
\delta n^{i}(\mathbf{x},t,\mathbf{v)}=n_{0}\frac{e\delta \phi }{T_{e}}\frac{1%
}{\omega }\left[ k_{y}V_{\ast e}-\omega k_{\perp }^{2}\rho _{s}^{2}\right] .
\label{idn}
\end{equation}

\bigskip

The quasineutrality condition leads to the dispersion equation

\begin{equation}
\frac{1}{\omega }\left[ k_{y}V_{\ast e}-\omega k_{\perp }^{2}\rho _{s}^{2}%
\right] =1+i\sqrt{\frac{\pi }{2}}\frac{\omega -k_{y}V_{\ast e}}{\left\vert
k_{z}\right\vert v_{Te}}
\end{equation}%
and to the well known solution for drift instability%
\begin{eqnarray}
\overline{\omega } &=&\frac{\overline{k}_{y}}{1+\overline{k}_{\perp }^{2}},
\label{lom} \\
\overline{\gamma } &=&\gamma _{0}\overline{\omega }\left( \overline{k}_{y}-%
\overline{\omega }\right) ,  \label{lgam}
\end{eqnarray}%
where $\overline{k}_{i}=k_{i}\rho _{s},$ $\overline{\omega }=\omega
L_{n}/c_{s},$ $\overline{\gamma }=\gamma L_{n}/c_{s},$ $\gamma _{0}=\sqrt{%
\pi /2}(c_{s}/L_{n})/\left\vert k_{z}\right\vert v_{Te}.$

As seen in the above equations, the drift instability is determined by the
nonadiabatic response of the electrons, which leads to $\gamma >0$ if $%
\omega <k_{y}V_{\ast e},$ condition ensured by the finite Larmor radius of
the ions. This conditions is fulfilled by all values of $k_{\perp }$ and
thus all the modes are unstable. The wave number domain of unstable modes is
very large, and the maximum growth rate is for $\omega =k_{y}V_{\ast e}/2,$\
which corresponds to $k_{\perp }\rho _{s}=1$. The growth rate is quadratic
in $k_{y}$ and thus the most unstable modes have the largest values of $%
k_{y} $ compatible with the above condition. These are the characteristics
of the linear (universal) drift instability on quiescent plasmas.

\bigskip

The solution in the zero Larmor radius limit is $\omega =k_{y}V_{\ast e},$ $%
\gamma =0,$ which represents the stable drift waves. For an arbitrary
initial condition $\phi _{0},$ this solution is 
\begin{equation}
\phi (x,y,z,t)=\phi _{0}(x,y\mathbf{-}V_{\ast e}t,z).  \label{s0}
\end{equation}%
\qquad Thus, the basic effect produced by the plasma when a potential
appears is to displace it with the diamagnetic velocity.

\bigskip

\subsection{Dispersion relation in turbulent plasma}

We consider a turbulent plasma with given statistical characteristics of the
stochastic potential $\phi _{b}(\mathbf{x},t).$ The potential is taken as
the zero order solution (\ref{s0}), obtained when the polarization drift is
neglected. It consists of the motion of the potential with the diamagnetic
velocity. The modification of potential shape and amplitude appear due to
polarization drift on a larger time scale of the order $1/\gamma .$ This
approximation is confirmed by the numerical simulations (see for instance 
\cite{Jenko} where the Eulerian correlation of the potential obtained from
numerical simulation of the trapped electron mode turbulence evidences the
motion of the potential with the diamagnetic velocity). The test mode
studies of turbulence are based on this time scale separation, which permits
a sequential approach. Starting from a potential that is a zero order
solution (\ref{s0}) it is possible to determine the frequency and the growth
rate of test modes as function of the statistical characteristics of the
potential. They provide information on the tendency in the evolution of the
potential, which is used to determine the test mode properties later in the
evolution, and so on.

The main statistical characteristics of the background turbulence are the
amplitude $\beta $ of the potential fluctuations, their correlation lengths $%
\lambda _{x},$ $\lambda _{y}$ and correlation time $\tau _{c}.$ These
parameters appear in the Eulerian correlation (EC) of the potential defined
by%
\begin{equation}
E(\mathbf{x},t)\equiv \left\langle \phi _{b}(\mathbf{x}^{\prime },t^{\prime
})\phi _{b}(\mathbf{x}^{\prime }+\mathbf{x},t^{\prime }+t)\right\rangle ,
\label{ECdef}
\end{equation}%
where $\left\langle {}\right\rangle $ is the statistical average or the
space average. This function is the Fourier transform of the spectrum. The
amplitude of the stochastic electric drift is $V=\sqrt{V_{x}^{2}+V_{y}^{2}},$
where $V_{x}=\beta /B\lambda _{y},$ $V_{y}=\beta /B\lambda _{x}.$ These
parameters define the time of flight (or the eddying time) $\tau
_{fl}=\lambda _{x}/V_{x}=\lambda _{x}\lambda _{y}B/\beta ,$ which is the
characteristic time for trajectory trapping.

The electron response to a perturbation with $\delta \phi $ of the
background potential $\phi _{b}$ is given by Eq. (\ref{edn}), as shown in
Section 2.1.

In order to determine ion response, we define the operator of derivation
along ion trajectories (where the parallel motion is neglected)%
\begin{equation}
\emph{O}^{i}\equiv \partial _{t}-\frac{\nabla \phi _{b}\times \mathbf{b}}{B}%
\cdot \nabla  \label{op}
\end{equation}%
The equation for the distribution function is%
\begin{equation}
\emph{O}^{i}f^{i}+f^{i}\mathbf{\nabla \cdot u}_{p}=0.
\end{equation}%
The distribution%
\begin{equation}
f_{0}^{i}=n_{0}(x)F_{M}^{i}\left( 1+\frac{e\phi _{b}(\mathbf{x-V}_{\ast e}t)%
}{T_{e}}\right)  \label{efi}
\end{equation}%
represents the approximate equilibrium because $\emph{O}^{i}f_{0}^{i}=0$ and
the term $\mathbf{\nabla \cdot u}_{p}\ll 1~$is of the order $V/\lambda
\Omega _{i}\tau _{c}.$ The divergence of the polarization drift has the
dimension of $t^{-1}$ and it introduces a characteristic time, which is of
the order $\tau _{p}=\tau _{c}\left( \Omega _{i}\tau _{fl}\right) $ thus
much larger than the correlation time of the potential. It means that for
time much smaller than this the remaining term is negligible and the ion
distribution function can be approximated by (\ref{efi}).

Perturbing the potential with $\delta \phi ,$ the operator is perturbed by 
\begin{equation*}
\delta \emph{O}^{i}=\left( 1/B\right) \left( -\mathbf{\nabla }\delta \phi
\times \mathbf{e}_{z}\right) \cdot \nabla
\end{equation*}
and a change of the distribution function appears $f^{i}=f_{0}^{i}+h.$ The
linearized equation in this perturbation is%
\begin{equation}
\emph{O}^{i}h+\delta \emph{O}^{i}f_{0}^{i}+h\mathbf{\nabla \cdot u}%
_{p}+f_{0}^{i}\mathbf{\nabla \cdot }\delta \mathbf{u}_{p}=0.
\end{equation}%
Since the background potential remains small compared with the kinetic
energy $e\phi _{b}/T_{e}<<1,$ the equilibrium distribution function in the
second and forth term can be approximated by $n_{0}(x)F_{M}^{i}$ 
\begin{equation}
\emph{O}^{i}h+h\mathbf{\nabla \cdot u}_{p}=-in_{0}(x)F_{M}^{i}\frac{e\delta
\phi }{T_{e}}\left( k_{y}V_{\ast e}-\omega \rho _{s}^{2}k_{\perp
}^{2}\right) .
\end{equation}%
This equations shows that the right side term is not influenced by the
turbulence (it is the same as for quiescent plasmas (\ref{gki})). The
effects of the background potential appear in the trajectories (in the
operator of derivation along trajectories) and in the second term which
accounts for the divergence of the polarization drift produced by the
turbulence.

The formal solution is%
\begin{equation}
h(\mathbf{x},v,t)=-n_{0}(x)F_{M}^{i}\frac{e\delta \phi }{T_{e}}\left(
k_{y}V_{\ast e}-\omega \rho _{s}^{2}k_{\perp }^{2}\right) \Pi ^{i}
\label{hi}
\end{equation}%
where the propagator is%
\begin{equation}
\Pi ^{i}=i\int_{-\infty }^{t}d\tau \exp \left[ i\mathbf{k\cdot }\left( 
\mathbf{x}(\tau )-\mathbf{x}\right) -i\omega \left( \tau -t\right) \right]
\exp \left[ -\int_{\tau }^{t}\mathbf{\nabla \cdot u}_{p}\left( \mathbf{x}%
(\tau )\right) \right]  \label{Pi}
\end{equation}%
and the integrals are along ion trajectories obtained from%
\begin{equation}
\frac{d\mathbf{x}(\tau )}{d\tau }=-\frac{\nabla \phi (\mathbf{x-V}_{\ast
e}t)\times \mathbf{e}_{z}}{B},  \label{traji}
\end{equation}%
with the condition at $\tau =t,$ $\mathbf{x}(t)=\mathbf{x.}$

The ion response (the propagator) is averaged over the stochastic
trajectories 
\begin{equation}
\overline{\Pi }^{i}=i\int_{-\infty }^{t}d\tau \ M(\tau ;t)\ \exp \left[
-i\omega \left( \tau -t\right) \right]  \label{piim}
\end{equation}%
where $M$ is the average 
\begin{equation}
M(\tau ;t)\equiv \left\langle \exp \left[ i\mathbf{k\cdot }\left( \mathbf{x}%
(\tau )-\mathbf{x}\right) +\frac{m_{i}}{eB^{2}}\int_{\tau }^{t}d\tau
^{\prime }\partial _{\tau ^{^{\prime }}}\Delta \phi \left( \mathbf{x}(\tau
^{\prime })\right) \right] \right\rangle .  \label{med}
\end{equation}%
The last term in the argument of the exponential accounts for the
compressibility effects in the background turbulence produced by the
polarization drift. This term has zero average, but as shown below, its
correlation with the displacements 
\begin{equation}
L_{i}(\tau )=\frac{m_{i}}{eB^{2}}\int_{\tau }^{t}d\tau ^{\prime \prime
}\int_{\tau }^{t}d\tau ^{\prime }\left\langle v_{i}\left( \mathbf{x}(\tau
^{\prime \prime })\right) ~\partial _{\tau ^{\prime }}\Delta \phi \left( 
\mathbf{x}(\tau ^{\prime })\right) \right\rangle .  \label{L}
\end{equation}%
is not zero and it can play a role in the evolution of the turbulence.

The ion density is%
\begin{equation}
\delta n^{i}(\mathbf{x},v,t)=-n_{0}(x)\frac{e\delta \phi }{T_{e}}\left(
k_{y}V_{\ast e}-\omega \rho _{s}^{2}k_{\perp }^{2}\right) \overline{\Pi }^{i}
\end{equation}%
and the dispersion relation for test modes on turbulent plasma is%
\begin{equation}
-\left( k_{y}V_{\ast e}-\omega \rho _{s}^{2}k_{\perp }^{2}\right) \overline{%
\Pi }^{i}=1+i\sqrt{\frac{\pi }{2}}\frac{\omega -k_{y}V_{\ast e}}{\left\vert
k_{z}\right\vert v_{Te}}  \label{dr}
\end{equation}%
since the electron density perturbation is the same as in quiescent plasma.

Thus, the effects of the background turbulence appear in the function of
time $M(\tau ;t)$ defined in Eq. (\ref{med}), which determines the ion
propagator. In the case of quiescent plasmas $M=1.$ This function imbeds all
the effects of the background turbulence. It depends implicitly on the
background potential through the statistical characteristics of the
trajectories (\ref{traji}), which determine the average, and explicitly
through the compressibility term $L_{i}$.

This function and its evolution is estimated in the next sections.

The polarization drift has an essential role. It destabilizes the drift
waves but it also has a more complex effect through the background
turbulence which is a weakly incompressible environment for the test modes
due to $\mathbf{u}_{p}.$ As shown below, the latter effect is important in
the strongly nonlinear regime.

\section{Test particles in turbulent plasma}

The $\mathbf{E}\times \mathbf{B}$ drift in turbulent plasmas can determine
trajectory trapping or eddying. It is permanent in the case of static
electric fields and appears due the Hamiltonian character of the motion with
the potential $\phi $ as conserved Hamiltonian. When this motion is weakly
perturbed by slow time variation of the potential or by other components of
the motion, trapping persists for finite time intervals determined by the
strength of the perturbation \cite{kraichnan}-\cite{mccomb}.

Semi-analytical statistical methods (the decorrelation trajectory method DTM 
\cite{V98} and the nested subensemble approach NSA \cite{VS04}) have been
developed for the study of test particle stochastic advection. A series of
studies (\cite{VS04} and the references there in) have shown that trapping
strongly influences the statistics of trajectories leading to memory
effects, quasi-coherent behavior and non-Gaussian distribution. The trapped
trajectories have quasi-coherent behavior and they form structures similar
to fluid vortices. The diffusion coefficients decrease due to trapping and
their scaling in the parameters of the stochastic field is modified \cite%
{V00}-\cite{V06}. Anomalous diffusion regimes and even sub-diffusion or
super-diffusion can appear due to trajectory trapping.

DTM and NSA reduce the problem of determining the statistical behavior of
the stochastic trajectories to the calculation of weighted averages of some
smooth, deterministic trajectories determined from the Eulerian correlation
(EC) of the stochastic field. These methods are in agreement with the
statistical consequences of the invariance of the potential. NSA is the
development of DTM as a systematic expansion that validates DTM and obtains
much more statistical information.

The main idea is to study the stochastic equation (\ref{traji}) in
subensembles of realizations of the stochastic field. First the whole set of
realizations $R$ is separated into subensembles $(S1)$ that contain all
realizations with given values of the potential and of the velocity in the
starting point of the trajectories. Then each subensemble $(S1)$ is
separated into subensembles $(S2)$ that correspond to fixed values of the
second derivatives of the potential. By continuing this procedure up to an
order $n$, a system of nested subensembles is constructed. The stochastic
(Eulerian) potential and velocity in a subensemble are Gaussian fields but
nonstationary and nonhomogeneous, with space- and time-dependent averages
and correlations. The correlations are zero in $\mathbf{x=0}$, $t=0$ and
increase with distance and time. The average potential and the average
velocity in a subensemble are determined by the Eulerian correlation of the
potential by conditional averages (see \cite{VS04} for details). The
stochastic equation (\ref{traji}) is studied in each highest-order
subensemble. Neglecting trajectory fluctuations, the average trajectory $%
\mathbf{X}(t;Sn)$ is obtained from an equation that has the structure of the
equation of motion (\ref{traji}), but with the stochastic potential replaced
by the subensemble average potential which is determined by the EC of the
potential. This approximation is acceptable because it is performed in the
subensemble where trajectories are similar as they are super-determined. In
addition to the necessary and sufficient initial condition $\mathbf{x}(t)=%
\mathbf{x,}$ supplementary initial conditions are imposed by the subensemble
definition. The strongest condition is the initial potential, which is a
conserved quantity in the static case and determines comparable trajectory
sizes in a subensemble. Moreover, the amplitude of the velocity fluctuations
in subensembles, the source of the trajectory fluctuations, is zero in the
starting point of the trajectories and reaches the value corresponding to
the whole set of realizations only asymptotically.

The statistical quantities are obtained as weighted averages of these
trajectories $\mathbf{X}(t;Sn)$ or of the functions of these trajectories.
The weighting factor is the probability that a realization belongs to the
subensemble $(Sn),$ and is analytically determined.

NSA is quickly convergent because the mixing of periodic trajectories, which
characterizes this nonlinear stochastic process, is directly described at
each order. Results obtained in the first order (DTM) for the diffusion
coefficient $D(t)$ are essentially not modified in the second order \cite%
{VS04}. Second-order NSA is important because it provides detailed
statistical information on trajectories, which contributes to the
understanding of the trapping process.

\bigskip

The EC of the potential is modelled according to the frequency and the
growth rates of drift modes in quiescent plasma (\ref{lom})-(\ref{lgam}).
This equations show that the maximum of $\gamma $ corresponds to $k_{\perp
}\rho _{s}=1,$ with larger $k_{y}$ (of the order of $\rho _{s}^{-1})$ and
smaller $k_{x}.$ The growth rate is zero for $k_{y}=0.$\ Thus, the spectrum
of the drift type turbulence is zero along the line $k_{y}=0,$ which means
that the integral of the EC along $y$ must be zero. It has two peaks at $%
k_{y}=\pm k_{0},k_{x}=0.$ A simple correlation with these properties that
also includes anisotropy is%
\begin{eqnarray}
E &=&\beta ^{2}e(x,y)f(y)  \label{rEC} \\
e(x,y) &=&\exp \left( -\frac{x^{2}}{2\delta }-\frac{y^{2}}{2}\right)  \notag
\\
f(y) &=&\cos k_{0}y-y\frac{\sin k_{0}y}{k_{0}}  \notag
\end{eqnarray}%
where $\beta $ is the amplitude of the fluctuations of the potential, $\
\delta =\lambda _{x}^{2}/\lambda _{y}^{2},$ the distances are normalized
with $\lambda =\lambda _{y}$ and $k_{0}$\ with $\lambda ^{-1}.$ This
correlation also accounts for the existing of dominant waves in the
stochastic potential. This correlation has negative values along $y$ (one
negative minimum for $k_{0}=0,$ and oscillatory behavior for large $k_{0}).$

Starting from the EC (\ref{rEC}), the statistics of ion trajectories is
determined, then the averaged ion propagator, the growth rate and the
frequency of the drift modes are estimated. These quantities show how the
correlation evolves.

The distribution of displacements obtained from Eqs. (\ref{traji}) strongly
depends on the ordering of the characteristic tines of the stochastic
process. The longest characteristic time is the parallel time of the ions $%
\tau _{\parallel i}=\lambda _{\parallel }/v_{Ti},$ which actually leads to
negligible parallel motion. The motion of the potential along $y$ with the
diamagnetic velocity defines the diamagnetic time $\tau _{\ast }=\lambda
_{y}/V_{\ast e}.$ The correlation time of the potential $\tau _{c}$ is the
characteristic time for the change of the shape of the potential $\tau
_{c}\cong \gamma ^{-1},$which is larger than the diamagnetic time ($\tau
_{c}>\tau _{\ast }).$ The time of flight (or eddying time) is defined by $%
\tau _{fl}=\lambda _{x}/V_{x}=B\lambda _{x}\lambda _{y}/\beta $. Trajectory
trapping appears when $\tau _{fl}$ is the smallest of these characteristic
times. The trapping parameter for drift turbulence is 
\begin{equation*}
K_{\ast }=\tau _{\ast }/\tau _{fl}=V_{y}/V_{\ast e}=\beta /(\lambda
_{x}V_{\ast e}B)
\end{equation*}%
and this process appears when $K_{\ast }>1.$

The average propagator (\ref{piim}) is calculated in the next sections using
simplified models that include the main characteristics of the probability
of displacements. It is so possible to capture the complicated nonlinear
effects of strong turbulence in rather simple analytical expressions. The
results can easily be improved by taking into account the statistics of test
particles obtained with the nested subensemble method. This significantly
more complicated approach that relies on numerical calculation of the
averages does not change qualitatively the results. We consider that the
simple analytical expressions derived in the next sections give a more clear
image on the complex nonlinear processes that appear in the drift turbulence
evolution beyond the quasi-linear stage.

\section{Ion diffusion and damping of large k modes}

For small amplitude of the background potential, the time of flight is
larger than the decorrelation time and trapping does not appear. The
smallest characteristic time that influence the ion motion corresponds to
the potential motion with the diamagnetic velocity, which travels over the
correlation length faster than the ions. The characteristic times ordering is%
\begin{equation}
\tau _{\parallel e}<\tau _{\ast }<\tau _{c}<\tau _{fl}<\tau _{\parallel i}.
\label{ord1}
\end{equation}%
This corresponds to small amplitude background turbulence with $\beta
/B\lambda _{x}\ll V_{\ast e},$ which does not produce trajectory trapping.
The EC of the turbulence is given by (\ref{rEC}) that corresponds to the
growth rates of the drift modes.

The diffusion coefficients determined for this quasilinear regime for a
frozen potential that moves with $V_{\ast e}$ and has the EC (\ref{rEC}) are
completely different from those obtained for an EC that is a decaying
function of the distance. In the latter case, the transport along the
average velocity is diffusive with $D_{y}=V_{y}^{2}\tau _{\ast
}=V_{y}^{2}\lambda _{y}/V_{\ast e}$ and the perpendicular transport is
subdiffusive with the decay of the diffusion coefficient given by the EC of
the potential $D_{x}(t)\sim \left\vert \partial _{y}E(V_{\ast
e}t)\right\vert /V_{\ast e}.$The EC (\ref{rEC}) determines similar
subdiffusion along $x$ and subdiffusion along $y$ also. The potential change
on the longer time scale $\tau _{c}$ leads to diffusive transport but with
very small diffusion coefficients on both directions, much smaller than for
a monotonically decaying potential. The dependence of the diffusion
coefficients on $\tau _{c}$ is different on the two directions: $D_{x}$
increases while $D_{y}$ decreases when $\tau _{c}$ increases. The
distribution of the displacements is Gaussian, as is the velocity.

The function (\ref{med}) calculated with the Gaussian distribution is given
by 
\begin{equation}
M(\tau ;t)=\exp \left[ -\frac{k_{i}^{2}\left\langle \left( x_{i}(\tau
)-x_{i}\right) ^{2}\right\rangle }{2}+ik_{i}L_{i}\right]  \label{M1}
\end{equation}%
where the quadratic term in the divergence of the polarization drift was
neglected. The quantity $L_{i}(\tau )$ defined in Eq. (\ref{L}) is a
Lagrangian correlation that has the dimension of a length and represents the
correlation of the compressibility term with ion trajectories in the
background turbulence.

The trajectories (the characteristics) have displacements until
decorrelation by potential motions that are small compared to the
correlation length and can be neglected. This strongly simplifies the
estimation of the compressibility term $L_{i}$ because the Lagrangian
correlation in (\ref{L}) can be approximated by the corresponding Eulerian
correlations. The correlation in $L_{x}$ is%
\begin{eqnarray*}
\left\langle v_{x}\left( \mathbf{x}(\tau ^{\prime \prime })\right) ~\partial
_{\tau ^{\prime }}\Delta \phi \left( \mathbf{x}(\tau ^{\prime })\right)
\right\rangle &\cong &\left\langle \partial _{x}\phi \left( \mathbf{x}-%
\mathbf{V}_{\ast e}\left( t-\tau ^{\prime \prime }\right) \right) ~\partial
_{\tau ^{\prime }}\Delta \phi \left( \mathbf{x}-\mathbf{V}_{\ast e}\left(
t-\tau ^{\prime }\right) \right) \right\rangle \\
&=&\frac{1}{V_{\ast e}}\partial _{\tau ^{\prime \prime }}\partial _{\tau
^{\prime }}\Delta E\left( \mathbf{V}_{\ast e}\left( \tau ^{\prime \prime
}-\tau ^{\prime }\right) \right)
\end{eqnarray*}%
which integrated over time gives%
\begin{equation}
L_{x}(\tau )=\frac{2}{BV_{\ast e}}\left( \Delta E\left( \mathbf{0}\right)
-\Delta E\left( \mathbf{V}_{\ast e}\left\vert t-\tau \right\vert \right)
\right) ,  \label{Lql}
\end{equation}%
which is finite because the Laplacean of the EC is symmetrical and finite in
zero. This function of time starts from zero at time $\tau =t$ and saturates
in a time $t-\tau $ of the order of$\ \tau _{\ast }$ at the value $2\Delta
E\left( \mathbf{0}\right) /BV_{\ast e}.\ $The other component $L_{y}$ is
zero because it contains one or three derivatives at $x$ for $x=0,$ which
are zero.

Thus in the quasilinear case%
\begin{eqnarray}
M &=&\exp \left[ -k_{i}^{2}D_{i}(t-\tau )+ik_{1}\frac{m_{i}}{eB^{3}}\frac{2}{%
V_{\ast e}}\Delta E\left( \mathbf{0}\right) \right]  \notag \\
&=&\exp \left[ -k_{i}^{2}D_{i}(t-\tau )-i2k_{1}\frac{V^{2}}{\Omega
_{i}V_{\ast e}}\right] ,
\end{eqnarray}%
which shows that the compressibility determines a constant term proportional
with the square of the amplitude of the stochastic velocity determined by
the background turbulence. This is very small due to $\Omega _{i}$ and to
the condition $V\ll V_{\ast e}$ imposed for this regime.

The propagator is%
\begin{equation}
\overline{\Pi }^{i}=-\exp \left[ -i2k_{1}\frac{V^{2}}{\Omega _{i}V_{\ast e}}%
\right] \frac{1}{\omega +ik_{i}^{2}D_{i}}  \label{Pi2}
\end{equation}%
and the ion density perturbation%
\begin{equation}
\delta n^{i}(\mathbf{x},v,t)=n_{0}(x)\frac{e\delta \phi }{T_{e}}A\frac{%
k_{y}V_{\ast e}-\omega \rho _{s}^{2}k_{\perp }^{2}}{\omega +ik_{i}^{2}D_{i}}
\label{drwnl}
\end{equation}%
where%
\begin{equation}
A=\exp \left[ -i2k_{1}\frac{V^{2}}{\Omega _{i}V_{\ast e}}\right] \approx
1-i2k_{1}\frac{V^{2}}{\Omega _{i}V_{\ast e}}.  \label{compr}
\end{equation}%
This factor produced by the polarization drift is very close to unity.

The dispersion relations is%
\begin{equation*}
\frac{k_{y}V_{\ast e}-\omega \rho _{s}^{2}k_{\perp }^{2}}{\omega
+ik_{i}^{2}D_{i}}A=1+i\sqrt{\frac{\pi }{2}}\frac{\omega -k_{y}V_{\ast e}}{%
\left\vert k_{z}\right\vert v_{Te}}
\end{equation*}%
and its solution is%
\begin{equation}
\overline{\omega }=\frac{\overline{k}_{y}}{1+\overline{k}_{\perp }^{2}},
\label{plqlom}
\end{equation}%
\begin{equation}
\overline{\gamma }=\gamma _{0}\overline{\omega }\left( \overline{k}_{y}-%
\overline{\omega }\right) -\overline{k}_{i}^{2}\overline{D}_{i}-2\overline{k}%
_{1}\overline{\omega }\frac{V^{2}}{\rho _{s}c_{s}},  \label{plqlgam}
\end{equation}%
where $\overline{D}_{i}=D_{i}/(\rho _{s}V_{\ast e}).$ Thus the frequency is
not influenced by the background turbulence when it has small amplitude. The
latter determines new terms in the growth rate.

The main effect is produced by ion trajectory diffusion in the background
turbulence, which determine the damping of the modes with large $k$ (the
second term in $\gamma ).$ This is the well known result of Dupree, which is
recovered here. An additional effect is obtained from the compressibility,
which determines the asymmetrical increase of the modes (contribution to the
growth of the modes with $k_{1}k_{2}<0$ and to the decay of those with $%
k_{1}k_{2}>0).$ This effect is small because the ratio of the third term to
the first term in Eq. (\ref{plqlom}) is of the order $\left( 1/\Omega
_{i}\tau _{\parallel e}\right) \left( V^{2}/V_{\ast e}^{2}\right) \ll 1$.

The frequency and the growth rate of the modes on turbulent plasma with
small amplitude show that the most unstable modes are, as in quiescent
plasma, those with $\omega \cong k_{y}V_{\ast e}/2,$ which have wave numbers
of the order of $\rho _{s}^{-1}$ ($\overline{k}_{\perp }^{2}\cong 1).$ Their
interaction with the background turbulence is weak, except for the large $k$
modes, which decay and eventually are damped due to ion diffusion.

This means that EC of the turbulence evolves in this quasilinear stage by
increasing its amplitude $\beta $ without major change of its shape.

\section{Trajectory structures and large scale correlations}

The increase of the amplitude $\beta $ of the stochastic potential
determines the change of the ordering of the characteristic times (\ref{ord1}%
). When $V$ becomes larger than the velocity of the potential $V_{\ast e}$
(or \ $\beta >V_{\ast e}\lambda _{y}B$), the time of flight is smaller that
the decorrelation time $\tau _{\ast }$ 
\begin{equation}
\tau _{\parallel }^{e}\ll \tau _{fl}<\tau _{\ast }\ll \tau _{c}\ll \tau
_{\parallel }^{i}.  \label{ord2}
\end{equation}%
In these conditions ion trapping or eddying appears. As we have shown this
strongly influences the statistics of trajectories. The distribution of the
trajectories is not more Gaussian due to trapped trajectories that form
quasi-coherent structures. At this stage the trapping is weak in the sense
that the fraction of trapped trajectories $n_{tr}$ is much smaller than the
fraction $n_{f}$ of free trajectories.

The probability of displacements $P(\mathbf{x},t)$ was obtained using the
nested subensemble method. It has a pronounced peaked shape compared to the
Gaussian probability. The contribution of the trapped particles to the
probability $P(\mathbf{x},t)$ at time $t$ appears in the narrow peak in $r=0$%
, which remains invariant after formation. The time invariance of this
central part of the probability is due to the vortical motion of the trapped
particles. It shows that vortical structures appear due to trapping. The
size of the vortical structure depends on the strength of the decorrelation
mechanism that releases the trapped particles. The free particles, for which
the motion is essentially radial, determines the larger distance part of the
probability, which continuously extends. The probability of displacements
until decorrelation ($\tau <\tau _{\ast })$ is modeled by

\begin{equation}
P(x,y,\tau )=n_{tr}G(\mathbf{x};\mathbf{S})+n_{f}G(\mathbf{x};\mathbf{S}%
^{\prime })  \label{prob}
\end{equation}%
where $G(\mathbf{x};\mathbf{S})$ is the 2-dimensional Gaussian distribution
with dispersion $\mathbf{S}=(S_{x},S_{y}).$ The first term describes the
trapped trajectories. We have considered for simplicity their distribution
as a Gaussian function but with small (fixed) dispersion that represents the
average size of the structures. The shape of this function does not change
much the estimations. The free trajectories are described by the second term
in Eq. (\ref{prob}). They have dispersion that grows linearly in time: $%
S_{i}^{\prime }\left( \tau \right) =S_{i}+2D_{i}\left( t-\tau \right) ,$ $%
i=x,y$. The initial value $S_{i}^{\prime }\left( t\right) =S_{i}$ is an
effect of trapping. It essentially means that the trajectories are spread
over a surface of the order of the size of the trajectory structures when
they are released by a decorrelation mechanism.

The compressibility term $L_{i}$ will be calculated in the next section. As
shown there, it is negligible in these conditions.

The average propagator is 
\begin{equation}
\overline{\Pi }^{i}=-\frac{1}{\omega +ik_{i}^{2}D_{i}}\mathcal{F},
\label{ap1}
\end{equation}%
where the factor $\mathcal{F}$\ is determined by the average size $\sqrt{%
S_{i}}$ of the trapped trajectory structures 
\begin{equation}
\mathcal{F\equiv }\exp \left( -\frac{1}{2}k_{i}^{2}S_{i}\right) .
\label{struct}
\end{equation}

The dispersion relation (\ref{dr}) is 
\begin{equation}
\frac{\left( k_{y}V_{\ast e}-\omega \rho _{s}^{2}k_{\perp }^{2}\right) }{%
\omega +ik_{i}^{2}D_{i}}\mathcal{F}=1+i\sqrt{\frac{\pi }{2}}\frac{\omega
-k_{y}V_{\ast e}}{\left\vert k_{z}\right\vert v_{Te}}  \label{rdstr}
\end{equation}%
with the solution%
\begin{equation}
\overline{\omega }=\frac{\mathcal{F}\overline{k}_{y}}{1+\mathcal{F}\overline{%
k}_{\perp }^{2}}  \label{omstr}
\end{equation}%
\begin{equation}
\overline{\gamma }=\gamma _{0}\overline{\omega }\left( \overline{k}_{y}-%
\overline{\omega }\right) -\overline{k}_{i}^{2}\overline{D}_{i}.
\label{gamstr}
\end{equation}%
Thus the effect of trajectory trapping appears in the frequency while the
growth rate is not modified (the small compressibility term was neglected).
The trajectory trapping determines the decrease of the frequency. The value
of $k_{\perp }$ corresponding to the maximum of $\gamma $ (at $\overline{%
\omega }=$ $\overline{k}_{y}/2)$ is displaced from values $\overline{k}%
_{\perp }\cong 1$\ to $\overline{k}_{\perp }\cong 1/\overline{s}$ where $%
\overline{s}=\sqrt{S_{x}+S_{y}}/\rho _{s}.$ For $\overline{s}\gg 1$ $\left(
S_{i}\gg \rho _{s}^{2}\right) ,$ the frequency (\ref{omstr}) is $\overline{%
\omega }\cong \mathcal{F}\overline{k}_{y},$ which shows that the finite
Larmor radius effects become negligible in the presence of trapping when the
size of the vortical structures is larger than $\rho _{s}.$

\begin{figure}[tbp]
\centerline{\includegraphics[height=6cm]{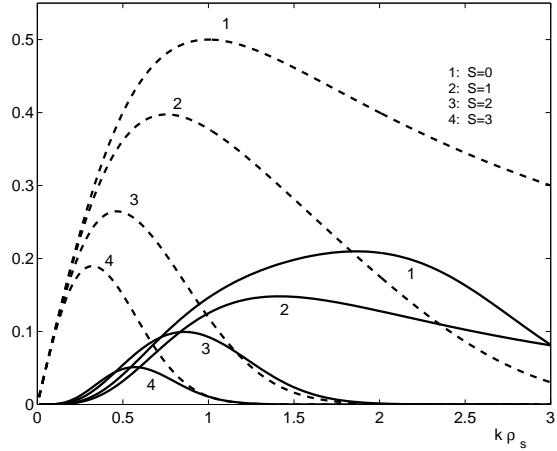}}
\caption{The frequencies (dashed curves) and the growth rates of drift modes
for several values of the size of trajectory structures }
\label{Figure 1}
\end{figure}

The growth rate of the drift modes for different values of the size $%
\overline{s}$ of the trajectory structures is shown in Figure 1. This shows
that the range of the wave numbers for the unstable test modes is displaced
toward small $k_{\perp }.$ This determines a further increase of the size of
trajectory structures. The diffusive term in (\ref{gamstr}) lead to the
damping of the modes with $\overline{k}_{\perp }\gg 1/\overline{s}.$\ 

This means that EC of the turbulence evolves in this weakly nonlinear stage
by increasing its amplitude $\beta $ and by increasing the correlation
length and the average wave number $k_{0}$ (\ref{rEC}). The vortical
structures of ion trajectories produced by trapping determine this processes.

\section{Ion flows and turbulence attenuation}

The evolution of the potential determines the increase of the fraction of
trapped ions. This determines an average flux of the trapped particles that
move with the potential $n_{tr}V_{\ast e}.$ As the $\mathbf{E}\times \mathbf{%
B}$ drift has zero divergence, the probability of the Lagrangian velocity is
time invariant, \textit{i. e.} it is the same with the probability of the
Eulerian velocity. The average Eulerian velocity is zero and thus the flux
of the trapped ions that move with the potential has to be compensated by a
flux of the free particles. These particles have an average motion in the
opposite direction with a velocity $V_{f}$ such that 
\begin{equation}
n_{tr}V_{\ast e}+n_{f}V_{f}=0.  \label{flows}
\end{equation}%
The NSA shows that the probability of the displacements splits in two
components that move in opposite direction. The peak of trapped ions moves
with the velocity $V_{\ast e}$ while the free ions move in the opposite
direction with a velocity that increases with the increase of the fraction
of trapped ions. The distribution of trapped ions is almost frozen while
that of free ions has a continuously growing width. Thus, opposite ion flows
are generated by the moving potential in the presence of trapping. A simple
approximation that includes the main features of the distribution is 
\begin{equation}
P(x,y,t)=n_{tr}G(x,y-V_{\ast
e}t;S_{x},S_{y})+n_{f}G(x,y-V_{f}t;S_{x}^{\prime },S_{y}^{\prime }).
\label{probg}
\end{equation}%
where two Gaussian functions were used as in Eq. (\ref{prob}) but taking
into account the ion flows (\ref{flows}). The average velocity of the free
ions is $V_{f}=-nV_{\ast e},~$where $n=n_{tr}/n_{f}.$ The separation of the
distribution and the existence of ion flows in drift type turbulence are
confirmed by numerical simulations \cite{Jenko}.

The compressibility term $L_{i}$ (\ref{L}) is determined for trapped and
free particles with the DTM. It can be written as%
\begin{equation*}
L_{x}=-\frac{V_{\ast e}}{\Omega _{i}B^{2}}\int_{\tau }^{t}d\tau ^{\prime
\prime }\int_{\tau }^{t}d\tau ^{\prime }\left\langle \partial _{y}\phi
\left( \mathbf{x}(\tau ^{\prime \prime })\right) ~\partial _{y}\Delta \phi
\left( \mathbf{x}(\tau ^{\prime })\right) \right\rangle .
\end{equation*}%
A nonzero value is obtained for $L_{x}$ essentially because the potential
and the Laplacean of the potential are correlated in a stochastic field: the
maxima of the potential statistically coincide with the minima of the
Laplacean. It follows that the derivatives of these functions along the same
directions are also correlated and the derivatives on different directions
are not. Indeed, one obtains from Eq. (\ref{rEC})%
\begin{eqnarray}
\left\langle \partial _{x}\phi \left( \mathbf{x}\right) ~\partial _{y}\Delta
\phi \left( \mathbf{x}\right) \right\rangle &=&-\partial _{x}\partial
_{y}\Delta E(\mathbf{0})=0  \label{c1} \\
\left\langle \partial _{y}\phi \left( \mathbf{x}\right) ~\partial _{y}\Delta
\phi \left( \mathbf{x}\right) \right\rangle &=&-\partial _{y}^{2}\Delta E(%
\mathbf{0})=-\frac{\beta ^{2}}{\lambda ^{4}}c,  \label{cx}
\end{eqnarray}%
where%
\begin{equation}
c=\left( \frac{k_{0}^{2}+3}{\delta }+15+10~k_{0}^{2}+k_{0}^{4}\right)
\label{cxn}
\end{equation}%
The corresponding Lagrangian correlations are determined with the DTM. The
component $L_{y}$ is zero for both trapped and free ions due to the symmetry
of the correlation (\ref{c1}). $L_{x}$ is different for free and trapped
trajectories. In the first case there is a relative motion of the free ions
and of the potential with the velocity $V_{\ast e}/n_{f}$ (obtained from (%
\ref{flows})), which determines rapid decorrelation. The estimation done for
the quasilinear turbulence also holds in these conditions and $L_{x}$ for
free ions is approximated by Eq. (\ref{Lql}) with $V_{\ast e}$ replaced by $%
V_{\ast e}/n_{f},~$and thus it is negligible. For the trapped trajectories, $%
L_{x}$ is much larger than in the quasilinear case. It can be determined
with the DTM, which shows that the following estimation holds. 
\begin{equation*}
L_{x}(\tau ,t)=-\frac{V_{\ast e}}{\Omega _{i}B^{2}}\int_{\tau }^{t}d\tau
^{\prime \prime }\int_{\tau }^{t}d\tau ^{\prime }C_{x}\left( \left\vert \tau
^{\prime \prime }-\tau ^{\prime }\right\vert \right) ,
\end{equation*}%
\begin{equation*}
C_{x}\left( \left\vert \tau ^{\prime \prime }-\tau ^{\prime }\right\vert
\right) =\left\langle \partial _{y}\phi \left( \mathbf{x}(\tau ^{\prime
\prime })-V_{\ast e}(t-\tau ^{\prime \prime })\right) ~\partial _{y}\Delta
\phi \left( \mathbf{x}(\tau ^{\prime })-V_{\ast e}(t-\tau ^{\prime })\right)
\right\rangle .
\end{equation*}%
The integral is calculated by changing the integration variable $\tau
^{\prime \prime }$ to $\theta =\tau ^{\prime \prime }-\tau ^{\prime }$%
\begin{equation*}
L_{x}(\tau ,t)=-\frac{2V_{\ast e}}{\Omega _{i}B^{2}}\int\limits_{0}^{t-\tau
}d\theta C_{x}\left( \left\vert \theta \right\vert \right) \left( t-\tau
-\theta \right) ,
\end{equation*}%
which integrated by parts gives%
\begin{equation*}
L_{x}(\tau ,t)=-\frac{2V_{\ast e}}{\Omega _{i}B^{2}}\int\limits_{0}^{t-\tau
}d\theta ~\emph{J}\left( \theta \right)
\end{equation*}%
where%
\begin{equation*}
\emph{J}\left( \theta \right) =\int\limits_{0}^{\theta }d\tau C_{x}\left(
\tau \right) .
\end{equation*}%
This function saturates after the decorrelation time $\tau _{d}$ at $\emph{J}%
\left( \theta \right) =C_{x}\left( 0\right) \tau _{d}.$ In these conditions
of strong turbulence, the decorrelation time for the $C_{x}$\ is, as for the
correlation of the stochastic $\mathbf{E\times B}$ velocity, the time of
flight $\tau _{fl}.$\ Thus%
\begin{equation}
L_{x}(\tau ,t)\cong a\left( t-\tau \right) ,\ \ \ a=2\partial _{y}^{2}\Delta
E(\mathbf{0})\frac{\tau _{fl}V_{\ast e}}{\Omega _{i}B^{2}},  \label{Lnl}
\end{equation}%
where $a$ has the dimension of a velocity. This type of Lagrangian
correlation corresponds to the decorrelation by mixing.

Thus, the compressibility of the background turbulence is correlated with
the displacements along $x$ and influences the propagator for the trapped
ions. The function $M$ (\ref{med}) that accounts for the effects of the
background turbulence is determined using Eqs. (\ref{probg}) and (\ref{Lnl}) 
\begin{eqnarray}
M &=&n_{f}\exp \left[ -\frac{k_{i}^{2}}{2}\left( S_{i}^{2}+2D_{i}(t-\tau
)\right) -ik_{y}V_{f}(t-\tau )\right] +  \notag \\
&&n_{tr}\exp \left[ -\frac{k_{i}^{2}}{2}S_{i}^{2}-ik_{y}V_{\ast e}(t-\tau
)+ik_{x}a(t-\tau )\right] .  \label{M2}
\end{eqnarray}%
The average propagator (\ref{piim}) becomes in these conditions%
\begin{equation}
\overline{\Pi }^{i}=-\mathcal{F}\left[ \frac{n_{f}}{\omega -k_{y}V_{f}}+%
\frac{n_{tr}}{\omega -k_{y}V_{\ast e}+k_{x}a}-ik_{i}^{2}D_{i}\frac{n_{f}}{%
\left( \omega -k_{y}V_{f}\right) ^{2}}\right]  \label{Pi3}
\end{equation}%
where the fact that the diffusion term is small was used.

The dispersion relation is%
\begin{eqnarray}
&&\mathcal{F}~\left( k_{y}V_{\ast e}-\omega k^{2}\rho _{s}^{2}\right) \left[ 
\frac{n_{f}}{\omega -k_{y}V_{f}}+\frac{n_{tr}}{\omega -k_{y}V_{\ast e}+k_{x}a%
}-i\frac{k_{i}^{2}D_{i}n_{f}}{\left( \omega -k_{y}V_{f}\right) ^{2}}\right] 
\notag \\
&=&1+i\sqrt{\frac{\pi }{2}}\frac{\omega -k_{y}V_{\ast e}}{\left\vert
k_{z}\right\vert v_{Te}}  \label{dlsnl}
\end{eqnarray}

The compressibility term $k_{x}a$ competes with the term $\omega
-k_{y}V_{\ast e}$ that is of the order $k_{y}V_{\ast e}$. Its effect is
measured by the ratio%
\begin{equation}
r=\frac{k_{x}a}{k_{y}V_{\ast e}}=2c\sqrt{\frac{\alpha }{k_{0}^{2}+3}}\frac{%
k_{x}}{k_{y}}\frac{\rho _{s}}{\lambda }\frac{V}{c_{s}}  \label{rap}
\end{equation}%
where $c$ is the normalized correlation (\ref{cxn})$.$ For a typical plasma
with $c_{s}=10^{6}m/s,$ $\beta =150V,$ $B=3T,$ $\lambda =10^{-2}m,$ $\rho
_{s}=2~10^{-3},$ one obtains $V/c_{s}=5~10^{-3}.$ The correlation $c\geq 20$%
, which leads to $r\cong 0.05$ for $k_{y}\cong k_{x}.$ It means that the
compressibility term can be neglected for modes with $k_{y}$ larger or of
the order of $k_{x}$ and that it is important for modes with $%
k_{y}\rightarrow 0.$

The solution of the dispersion relation is determined for the two cases: $%
k_{y}\gtrsim k_{x}$ \ that corresponds to drift modes and $k_{y}=0,$ which
are modes with different characteristics called zonal flow modes.

\bigskip

\subsection{Drift modes}

\bigskip Neglecting the compressibility term $\left( a=0\right) $, the real
part of the dispersion relation (\ref{dlsnl}) is%
\begin{equation}
\mathcal{F}\left( \overline{k}_{y}-\overline{\omega }_{d}\overline{k}_{\perp
}^{2}\right) \left[ \overline{\omega }_{d}+\left( n-1\right) \overline{k}_{y}%
\right] =\left( \overline{\omega }_{d}+n\overline{k}_{y}\right) \left( 
\overline{\omega }_{d}-\overline{k}_{y}\right) ,  \label{rp}
\end{equation}%
where the suscript was introduced in $\overline{\omega }_{d}$ to account for
drift type modes. Taking $\overline{\omega }_{d}=\overline{\omega }_{0}+%
\overline{\delta }(n,k),$ where $\overline{\omega }_{0}=\mathcal{F}\overline{%
k}_{y}/\left( 1+\mathcal{F}\overline{k}_{\perp }^{2}\right) $ is the
frequency (\ref{omstr}) in the absence of ion flows $(n=0),$ the equation
becomes%
\begin{equation}
\overline{\delta }\left[ \overline{\delta }+\overline{\omega }_{0}+\left(
n-1\right) \overline{k}_{y}\right] =\frac{n\overline{k}_{y}^{2}}{1+\mathcal{F%
}\overline{k}_{\perp }^{2}},  \label{rdelta}
\end{equation}%
where Doppler shift with the velocity of the trapped and free ions appears
in the parenthesis in the left side term, $\omega
_{0}-k_{y}V_{f}-k_{y}V_{\ast e}=k_{y}V_{\ast e}\left[ \overline{\omega }%
_{0}+\left( n-1\right) \overline{k}_{y}\right] .$ The solutions are the two
intersections of the parabola in the left hand side of Eq. (\ref{rdelta})
with the horizontal line at the level $n\overline{k}_{y}^{2}/\left( 1+%
\mathcal{F}\overline{k}_{\perp }^{2}\right) .$ The condition $\overline{%
\delta }=0$ for $n=0$ \ chooses one of the two solutions, namely on the
branch of the parabola that passes through the origin. This leads to $%
\overline{\delta }<0$\ if $\overline{\omega }_{0}+\left( n-1\right) 
\overline{k}_{y}<0$ and $\overline{\delta }>0$\ if $\overline{\omega }%
_{0}+\left( n-1\right) \overline{k}_{y}>0.$ The absolute value of $\overline{%
\delta }$ is an increasing function of $n$\ in both cases, and thus, as $n$
increases, the decrease of $\overline{\omega }_{d}$ is produced by the ion
flows in the first case and the increase of $\overline{\omega }_{d}$ in the
second case. The solution of the dispersion relation (\ref{rp}) is%
\begin{equation*}
\overline{\omega }_{d}=\frac{1}{2}\left[ \overline{\omega }_{0}+\left(
1-n\right) \overline{k}_{y}+sg\sqrt{\left[ \overline{\omega }_{0}+\left(
n-1\right) \overline{k}_{y}\right] ^{2}+\frac{4n\overline{k}_{y}^{2}}{1+%
\mathcal{F}\overline{k}_{\perp }^{2}}}\right] ,
\end{equation*}%
\ where $sg=sign\left[ \overline{\omega }_{0}+\left( n-1\right) \overline{k}%
_{y}\right].$ A jump appears in the frequency at the value of $k_{d}$
determined by the equation $\overline{\omega }_{0}=\left( 1-n\right) 
\overline{k}_{y},$ which has solution if $n<1.$ $k_{d}$\ is an increasing
function of $n.$In the limit of large $k,$ $\overline{\omega }%
_{d}\rightarrow -n\overline{k}_{y}$ if $n<1$ and $\overline{\omega }%
_{d}\rightarrow \overline{k}_{y}$\ if $n>1.$ This means that the asymptotic
phase velocity is in the ion diamagnetic direction when $n<1$ and in the
electron diamagnetic direction when trapping is stronger such that $n>1.$ \ 

The imaginary part (obtained using the condition $\overline{\gamma }_{d}\ll 
\overline{\omega }_{d})$ is 
\begin{eqnarray*}
&&\overline{\gamma }_{d}\mathcal{F}\left[ \left( 1-\overline{\omega }_{d}\ 
\overline{k}_{\perp }^{2}\right) f_{+}+\overline{k}_{\perp }^{2}\left( \frac{%
n_{f}}{\overline{\omega }_{d}+n\overline{k}_{y}}+\frac{n_{tr}}{\overline{%
\omega }_{d}-\overline{k}_{y}}\right) \right] \\
&=&\gamma _{0}\left( \overline{k}_{y}-\overline{\omega }_{d}\right) -n_{f}%
\overline{k}_{i}^{2}\overline{D}_{i}\frac{\mathcal{F}\left( 1-\overline{%
\omega }_{d}\overline{k}_{\perp }^{2}\right) }{\left( \overline{\omega }%
_{d}+n\overline{k}_{y}\right) ^{2}}
\end{eqnarray*}%
where 
\begin{equation*}
f_{+}=\left( \frac{n_{f}}{\left( \overline{\omega }_{d}+n\overline{k}%
_{y}\right) ^{2}}+\frac{n_{tr}}{\left( \overline{\omega }_{d}-\overline{k}%
_{y}\right) ^{2}}\right) >0.
\end{equation*}%
Using (\ref{rp}), the solution can be written as%
\begin{equation}
\overline{\gamma }_{d}=\frac{\gamma _{0}\left( \overline{\omega }_{d}+n%
\overline{k}_{y}\right) \left[ \left( 1-n\right) \overline{k}_{y}-\overline{%
\omega }_{d}\right] -n_{f}\overline{k}_{i}^{2}\overline{D}_{i}}{\left[
\left( 1-n\right) \overline{k}_{y}-\overline{\omega }_{d}\right] ^{2}\left(
1+\mathcal{F}\overline{k}_{\perp }^{2}\right) +n\overline{k}_{y}^{2}}\left( 
\overline{k}_{y}-\overline{\omega }_{d}\right) ^{2}.  \label{gammadr}
\end{equation}

The growth rates are presented in Figure 2a. The jump in the frequency
corresponds to a jump in the growth rate from negative (at small $k$) to
positive values. Thus the large wave-length modes are stabilized by the ion
flows, while the small wave-length modes still grow. This leads to the
increase of the amplitude of turbulence accompanied by the decrease of its
correlation length. Both effects contribute to the increase of the ratio $%
\tau _{\ast }/\tau _{fl}=\beta /(V_{\ast e}\lambda _{y}B),$ which leads to
the increase of $n$. Consequently, the increase of the amplitude of the
turbulence continues as well as the damping of small $k$ modes and the
decrease of the correlation length. The process is stopped when $n>1$
because the growth rate becomes negative for the whole wave number range.
This determines the decrease of $n$\ and the attenuation of the ion flows.

\begin{center}
\begin{figure}[!htb]
\centering
\begin{minipage}[t]{0.4\linewidth}
    \centering
    \includegraphics[height=5cm]{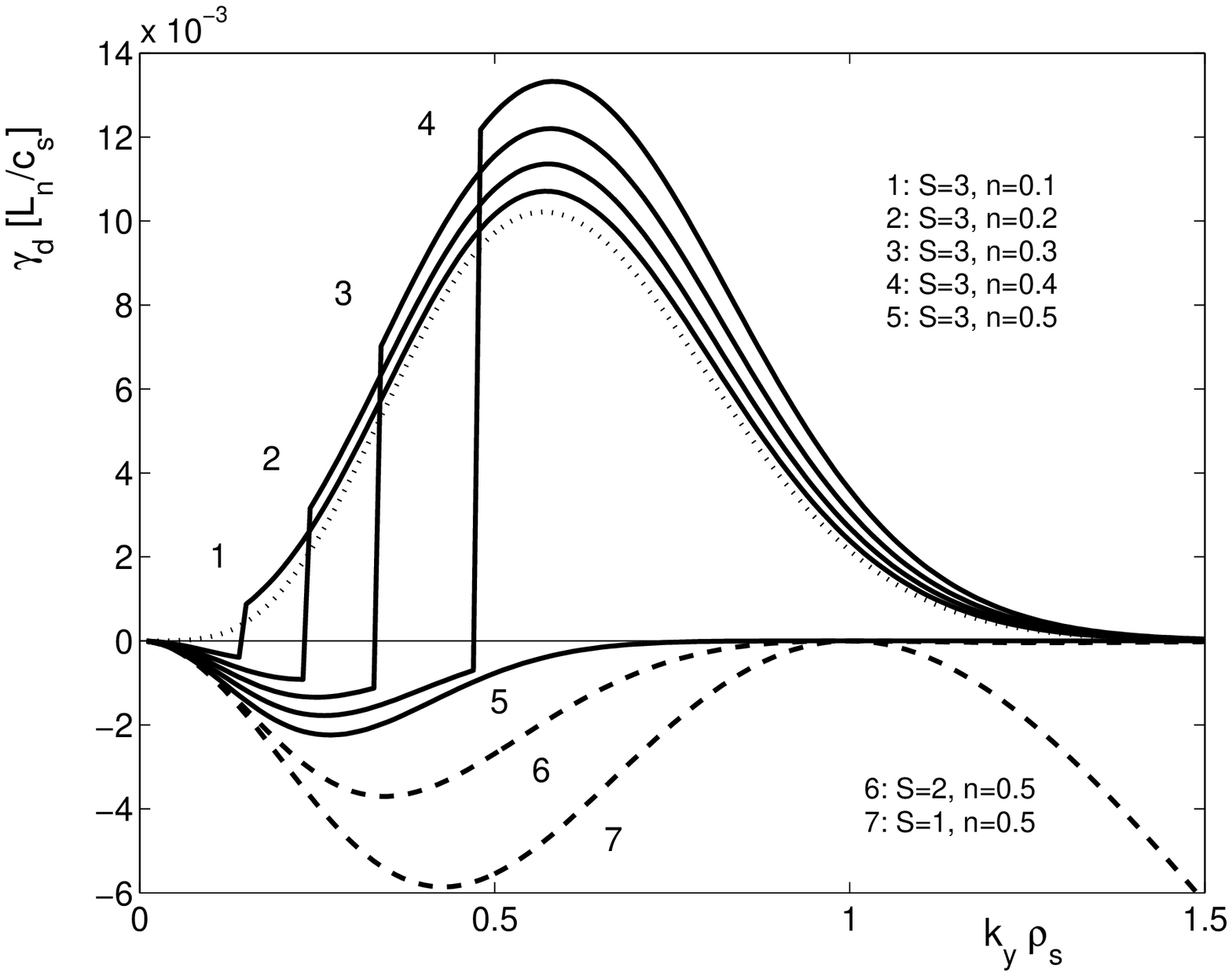}
   \end{minipage}
\hspace{0.1\textwidth}  
\begin{minipage}[t]{0.4\linewidth}
    \centering
    \includegraphics[height=5cm]{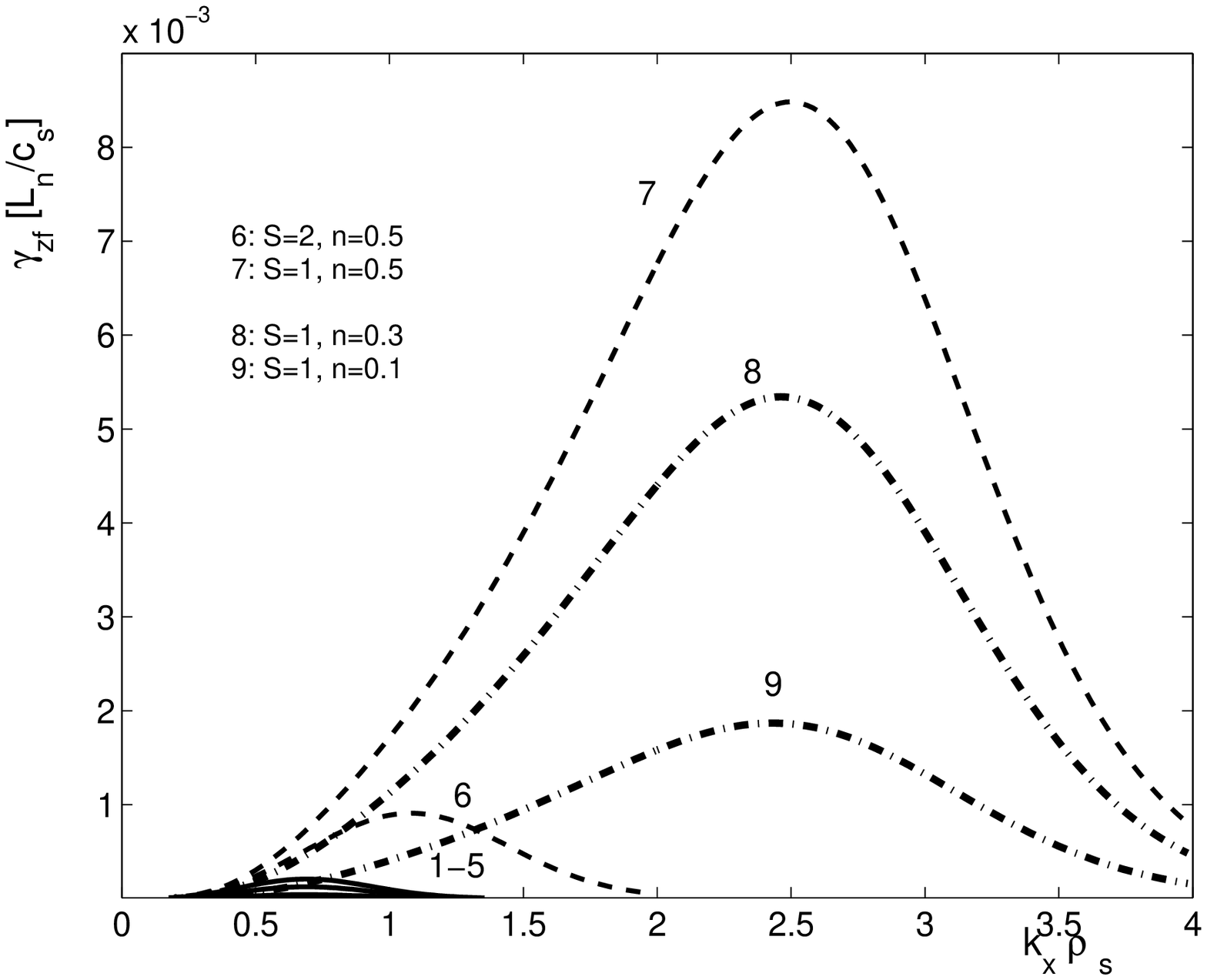}
   \end{minipage}
\caption{Growth rate dependence on the fraction of trapped ions $n$ and on
the size $S$ of trajectory structures for drift modes (left panel) and zonal
flow modes (right panel). The curves with the same label correspond to the
same parameters. }
\label{Figure 2}
\end{figure}
\end{center}

\subsection{Zonal flow modes ($k_{y}=0)$}

The compressibility term becomes dominant for $k_{y}=0.$ As shown below it
determines unstable modes completely different of the drift modes: with $%
k_{y}=0,$ $k_{x}\neq 0$ and very small frequency $\omega _{zf}$, much
smaller than the diamagnetic frequency. These are zonal flow modes.

The dispersion relation for $k_{y}=0$ is%
\begin{equation*}
-\omega _{zf}\mathcal{F}k_{x}^{2}\rho _{s}^{2}\left[ \frac{n_{f}}{\omega
_{zf}+ik_{i}^{2}D_{i}}+\frac{n_{tr}}{\omega _{zf}+k_{x}a}\right] =1+i\sqrt{%
\frac{\pi }{2}}\frac{\omega _{zf}}{\left\vert k_{z}\right\vert v_{Te}},
\end{equation*}%
For $\gamma _{zf},$ $k_{i}^{2}D_{i}\ll \omega _{zf},$ the real part is%
\begin{equation*}
-\omega _{zf}\mathcal{F}k_{x}^{2}\rho _{s}^{2}\left( \omega
_{zf}+n_{f}k_{x}a\right) =\omega _{zf}\left( \omega _{zf}+k_{x}a\right)
\end{equation*}%
with the solution%
\begin{equation}
\overline{\omega }_{zf}=-\overline{k}_{x}\overline{a}\frac{1+\mathcal{F}%
\overline{k}_{x}^{2}n_{f}}{1+\mathcal{F}\overline{k}_{x}^{2}}
\end{equation}%
where 
\begin{equation*}
\overline{a}=\frac{a}{V_{\ast e}}=2c\sqrt{\frac{\alpha }{k_{0}^{2}+3}}\frac{%
\rho _{s}}{\lambda }\frac{V}{c_{s}}
\end{equation*}%
The imaginary part is%
\begin{eqnarray*}
&&\gamma _{zf}\mathcal{F}k_{x}^{2}\rho _{s}^{2}\left[ \omega _{zf}\left( 
\frac{n_{f}}{\omega _{zf}^{2}}+\frac{n_{tr}}{\left( \omega
_{zf}+k_{x}a\right) ^{2}}\right) -\frac{n_{f}}{\omega _{zf}}-\frac{n_{tr}}{%
\omega _{zf}+k_{x}a}\right] \\
&=&\sqrt{\frac{\pi }{2}}\tau _{\parallel e}\omega _{zf}-k_{x}^{4}D_{x}%
\mathcal{F}\rho _{s}^{2}\frac{n_{f}}{\omega _{zf}}
\end{eqnarray*}%
with the solution 
\begin{equation}
\overline{\gamma }_{zf}=\left[ \gamma _{0}\left( \overline{k}_{x}\overline{a}%
\right) ^{2}\frac{\left( 1+n_{f}\mathcal{F}\overline{k}_{x}^{2}\right) ^{2}}{%
\left( 1+\mathcal{F}\overline{k}_{x}^{2}\right) ^{2}}-n_{f}\mathcal{F}%
\overline{k}_{x}^{4}\overline{D}_{x}\right] \frac{n_{tr}\mathcal{F}\overline{%
k}_{x}^{2}}{\left( 1+n_{f}\mathcal{F}\overline{k}_{x}^{2}\right) \left( 1+%
\mathcal{F}\overline{k}_{x}^{2}\right) }
\end{equation}%
These unstable zonal flow modes are consequence of trapping combined with
the polarization drift. One can see that when trapping is negligible ($%
n_{tr}\cong 0),$ $\omega _{zf}=-k_{x}a$ and $\gamma _{zf}=0,$ and that $%
\omega _{zf},\gamma _{zf}=0$ for $\overline{a}=0.$ As seen in Figure 2b, the
growth rate of zonal flows is positive and it increases with the increase of 
$n_{tr}$ and with the decrease of the size of the trajectory structures.

Thus we have shown that zonal flow modes are generated due to ion flows.
They have positive growth rates and small frequencies (typically ten times
smaller than the diamagnetic frequency).

\bigskip

In conclusion, the ion flows produced by ion trapping in the moving
potential determine two parallel effects: nonlinear damping of drift modes
and generation of zonal flow modes. There is no causality relation between
zonal flow modes and drift turbulence damping. Both effects are generated
selfconsistently in the nonlinear evolution of drift turbulence. The
influence produced by the zonal flows on the drift type modes is only
indirect, through the diffusive damping. Zonal flows modify the
configuration of the potential and consequently the correlation of the
turbulence by introducing components with $k_{y}\cong 0$ in the spectrum,
which change the shape of the EC. The integral of the EC along $y$ becomes
finite. This determines a rather strong increase of the diffusion
coefficient $D_{y},$ which contribute to the decay of the drift type
turbulence. Zonal flow modes also modify the diffusion along $x$ but in the
opposite sense: a strong decrease appears due to the decrease of $\lambda
_{x}$ produced by the zonal flow component of the turbulence. The drift
turbulence decay determines the decrease of $n_{tr},$\ which reduces the
growth rate of zonal flow modes.

There is time correlation between the maximum growth rate of zonal flow
modes and the damping of the drift modes, as observed in experiments and in
the numerical simulations. This time correlation can be deduced from Figures
2 (a, b), which show the growth rates of the drift and zonal flow modes for
the same set of parameters $n,\ S.$ When the drift turbulence amplitude
increases determining the increase of the fraction $n$ of trapped ions
(curves 1-5) the growth rate of the zonal flows is very small. When the
growth rate of the drift modes becomes negative and the size of the ion
trajectory structures decreases (curves 6, 7 in Figure 2) the growth rate of
the zonal flow modes strongly increases and, at small $S,$ it becomes
comparable with that of the drift modes. The attenuation of the drift
turbulence that leads to the decrease of $n$ determines the decrease of the
growth rate of the zonal flow modes (curves 8, 9 in Figure 2 b).

We note that the effect of ion diffusion on the zonal flow modes and of
drift modes is different. Ion diffusion determines the damping of the drift
modes with large $k.$ This effect is well known and appears on the whole
evolution of the drift modes, from the quasilinear to the strongly nonlinear
stage characterized by the existence of ion flows. In the case of zonal flow
modes, Eq. (\ref{gamzf}) shows that the effect of ion diffusion on large $%
k_{x}$ modes is negligible due to the factor $\mathcal{F}$ that goes to
zero. The diffusion has strongest effect on the modes with $k_{x}\cong \sqrt{%
1/S_{x}},$ and it can even determine negative $\gamma $ at small $k_{x}.$

The evolution of the turbulence in this strongly nonlinear stage is rather
complex. The ion flows determine first the increase of the amplitude of the
turbulence and the decrease of the correlation length $\lambda _{y}$, then a
major modification of the EC shape appears due to generation of zonal flow
modes and eventually the turbulence is strongly attenuated. The new
component that accounts for the zonal flow modes adds to the EC a function
with a slow decay along $y$ ($k_{y}\cong 0),$ with zero integral along $x$
(because $\gamma =0$ for $k_{x}=0)$\ and with amplitude growing with the
increase of $n.$ It essentially determines in the total EC the increase of $%
\lambda _{y}$ and of the amplitude $\beta $ and the decrease of $\lambda
_{x}.$

\section{Summary and conclusions}

Test modes on turbulent plasmas were studied taking into account the process
of ion trajectory trapping in the structure of the background potential. The
case of drift turbulence was considered, and the frequency $\omega $ and the
growth rate $\gamma $ were determined as functions of the statistical
properties of the background turbulence. The main characteristics of the
evolution of the turbulence were deduced from $\gamma $ and $\omega .$

A different physical perspective on the nonlinear evolution of drift
turbulence is obtained.

The main role is played by the trapping of the ions in the stochastic
potential that moves with the diamagnetic velocity. The fraction of trapped
ions is a function of $K_{\ast }=\beta /(\lambda _{x}V_{\ast e}B).$ Trapping
appears for $K_{\ast }>1$ and the fraction of trapped trajectories $n_{tr}$
increases with the increase of this parameter. Trapped trajectories form
quasi-coherent structures, which determine non-Gaussian distribution of the
displacements. At larger amplitudes, the moving potential determines ion
local flows: the trapped ions move with the potential while the other ions
drift in the opposite direction. These opposite (zonal) flows compensate
such that the average ion velocity is zero, but they determine the spliting
of the distribution of displacements.

The first effect of trapping in turbulence evolution appears when the
fraction of trapped ions is small and the ion flows are negligible. The
trapped ions determine the evolution of the turbulence toward large wave
lengths (the inverse cascade).

The influence of the ion flows produced by the moving potential appears
later in the evolution of the turbulence when the fraction of trapped ions
is comparable whith that of free ions. The ion flows determine first the
damping of the small $k_{\perp }$ drift modes and eventually the damping of
the drift modes with any $k_{\perp }.$ In the same time, trapping in
connection with the compressibility produced by the polarization drift in
the background turbulence determine transitory zonal flow modes (with $%
k_{y}=0$ and $\omega \ll k_{y}V_{\ast e}).$ Thus, in this perspective, there
is no causality connection between the damping of the drift turbulence and
the zonal flow modes. Both processes are produced by ion trapping in the
moving potential.

During the decay of the turbulence trapping is maintained because the
decrease of the amplitude is compensated by decrease of the correlation
length, leading to small variations of the parameter $K_{\ast }$. The
damping is first produced for the small wave numbers and thus there is a
transitory domination by large wave numbers in the spectrum. This determines
a hysteresis process: the growth and the decay of the turbulence is produced
on different paths. Large scales are generated at the development (increase)
of the turbulence and the correlation length increases. This determines only
a slow increase of the parameter $K_{\ast }$ and thus a slow increase of the
number of trapped trajectories $n_{tr}.$Then, the ion flows become important
and produce further increase of the larger part of the wave number spectrum
accompanied by the decay of the small $k$ modes. Thus, the amplitude of the
potential increases and the correlation length decrease. This determines a
much faster increase of $K_{\ast }$ and of the relative number of trapped
ions. When the latter is $n_{tr}=0.5,$ the growth rates of all drift modes
are negative and the amplitude of the turbulence decays leading to the
decrease of $n_{tr}$. This process continues until the ion flows become
negligible ($n_{tr}\ll n_{f}).$ A closed evolution curve in the $\left(
\beta ,~\lambda \right) $ space is described by the turbulence, which
remains in the nonlinear stage characterized by trapping and oscillates
between weak and strong trapping.

The zonal flows appear due to ion flows as well as turbulence nonlinear
damping. The damping process is not determined by the zonal flows. There is
however an influence produced by the zonal flows on the drift type modes. It
is not direct but through the diffusive damping. Zonal flows modify the
correlation of the turbulence by introducing components with $k_{y}=0$ in
the spectrum. This changes the shape of the EC, which has no more zero
integral along the $y$ direction. This determines the increase of the
diffusion $D_{y},$ which contribute to the decay drift type turbulence but
not of the zonal flow modes.

The diffusion coefficient along $x$\ axis also oscillates during the
evolution of the turbulence. Its maximum is correlated with the period when
the correlation lengths along $x$ is large, which appears before the
development of the ion flows.

The zonal flow modes increase and drift turbulence decay appear to be
temporally correlated, as observed in experiments and simulations. The
characteristic time $\Delta t$ for turbulence and transport oscillations can
be estimated as the inverse of the growth rates, which are of the order of $%
5\ 10^{-3}c_{s}/L_{n}.$ One obtains for $L_{n}/R_{0}=5,$ $\Delta t=40\
R_{0}/c_{s},$ in agreement with the numerical simulations.

\bigskip

%%\begin{acknowledgement}
This work was supported by CNCSIS-UEFISCDI, project number PNII - IDEI 1104/2008.
%%\end{acknowledgement}


\begin{thebibliography}{99}
\bibitem{K02} Krommes J. A., \emph{Phys. Reports }\textbf{360} (2002) 1.

\bibitem{Horton} Horton W, \emph{Rev. Modern Phys.} \textbf{71} (1999) 735.

\bibitem{garbet} Garbet X, Idomura Y, Villard L, Watanabe T H, \emph{Nuclear
Fusion} \textbf{50 }(2010) 043002.

\bibitem{tynan} Tynan G. R., Fujisawa A., McKee G., \emph{Plasma Phys.
Control. Fusion} \textbf{51} (2009) 113001.

\bibitem{Tzf} Terry PW, Rev. Mod. Phys. 72, 109 (2000)

\bibitem{Diamond} Diamond P. H., Itoh S.-I., Itoh K., Hahm T. S., \emph{%
Plasma Phys. Control. Fusion} \textbf{47} (2005) R35-R161.

\bibitem{V98} Vlad M., Spineanu F., Misguich J.H., Balescu R., $\emph{%
Phys.Rev.E}$ \textbf{58} (1998) 7359.

\bibitem{VS04} Vlad M. and Spineanu F., \emph{Phys. Rev. E} \textbf{70 }%
(2004) 056304.

\bibitem{D66} Dupree T. H., \emph{Phys. Fluids} \textbf{9} (1966) 1773.

\bibitem{D72} Dupree T. H., \emph{Phys. Fluids} \textbf{15} (1972) 334.

\bibitem{SV88} Spineanu F., Vlad M., \emph{Phys. Letters A} \textbf{133}
(1988) 319.

\bibitem{VSM} Vlad M., Spineanu F., Misguich J.H., \emph{Plasma Phys.
Control. Fusion} \textbf{36} (1994) 95.

\bibitem{GR} Goldstone R. J. and Rutherford P. H., \textit{Introduction to
Plasma Physics}, Institute of Physics Publishing, Bristol and Philadelphia,
1995.

\bibitem{Jenko} Hauff T, Jenko F, \emph{Phys. Plasmas }\textbf{14} (2007)
092301.

\bibitem{kraichnan} R. H. Kraichnan, \emph{Phys. Fluids} \textbf{19}, 22
(1970).

\bibitem{Bcarte} Balescu R., \textit{Aspects of Anomalous Transport in
Plasmas}, Institute of Physics Publishing (IoP), Bristol and Philadelphia,
2005.

\bibitem{mccomb} W.\ D.\ McComb, \textit{The Physics of Fluid Turbulence}
(Clarendon, Oxford,\ 1990).

\bibitem{V00} Vlad M., Spineanu F., Misguich J.\ H. and Balescu R., \emph{%
Phys. Rev. E} \ \textbf{61} (2000) 3023.

\bibitem{V01} Vlad M., Spineanu F., Misguich J.\ H. and Balescu R., \emph{%
Phys. Rev. E } \textbf{63} (2001) 066304.

\bibitem{V02} Vlad M., Spineanu F., Misguich J.\ H. and Balescu R., \emph{%
Nuclear Fusion }\textbf{42} (2002) 157.

\bibitem{V04} Vlad M., Spineanu F., Misguich J.\ H., Reusse J.-D., Balescu
R., Itoh K., Itoh S.-I., \emph{Plasma Phys. Control. Fusion }\textbf{46}
(2004) 1051.

\bibitem{P07} Petrisor I., Negrea M., Weyssow B., \emph{Physica Scripta }%
\textbf{75} (2007) 1.

\bibitem{V03} Vlad M., Spineanu F., Misguich J.H., Balescu R., \emph{Phys.
Rev. E }\textbf{67} (2003) 026406.

\bibitem{NS} Neuer M., Spatschek K. H., \emph{Phys. Rev. E } \textbf{74 (}%
2006\textbf{) }036401.

\bibitem{V06} Vlad M., Spineanu F., Benkadda S., \emph{Phys. Rev. Letters }%
\textbf{96} (2006) 085001.
\end{thebibliography}
\end{document}